\documentclass[prl,twocolumn,showpacs,superscriptaddress,amsfonts,amsmath,
floatfix]{revtex4}
\usepackage{graphicx}
\usepackage{psfrag}
\newcommand{\Journal}[4]{#1 {\bf #2}, #3 (#4)}
\newcommand{\PR}{Phys. Rev.}
\newcommand{\PRL}{Phys. Rev. Lett.}
\newcommand{\PRA}{Phys. Rev. A}

\newcommand{\JMP}{J. Math. Phys.}

\newcommand{\Science}{Science}
\newcommand{\PLA}{Phys. Lett. A}
\begin{document}
\title {Soluble Models of Strongly Interacting Ultracold 
Gas Mixtures in Tight Waveguides}
\author{M. D. Girardeau}
\email{girardeau@optics.arizona.edu}
\affiliation{College of Optical Sciences, University of Arizona, 
Tucson, AZ 85721, USA}
\author{A. Minguzzi}
\email{anna.minguzzi@grenoble.cnrs.fr}
\affiliation{Laboratoire de Physique et Mod\'elisation des Mileux Condens\'es, 
C.N.R.S., B.P. 166, 38042 Grenoble, France}
\date{\today}

\begin{abstract}
A Fermi-Bose mapping method is used to determine the exact
ground states of several models of mixtures of strongly 
interacting ultracold gases in tight waveguides, which are generalizations
of the Tonks-Girardeau (TG) gas (1D Bose gas with point hard cores) 
and fermionic Tonks-Girardeau (FTG) gas (1D spin-aligned
Fermi gas with infinitely strong zero-range attractions). 
We detail the case of  a Bose-Fermi mixture with TG boson-boson (BB)
and boson-fermion (BF) interactions. Exact results are given for density 
profiles in a harmonic trap, single-particle density matrices, momentum 
distibutions, and density-density correlations.  Since the ground state is 
highly degenerate, we analyze the splitting of the ground manifold for 
large but finite  BB and BF repulsions. 
\end{abstract}
\maketitle
Due to the rapidly increasing sophistication of experimental techniques
for probing ultracold gases, theoretical emphasis has shifted from effective 
field approaches 
to more refined methods capable of dealing with correlations, and experiments
measuring such correlations have been carried out.
In ultracold gases confined in de Broglie waveguides with transverse trapping 
so tight that the atomic dynamics is essentially one-dimensional \cite{Ols98}, 
with confinement-induced resonances \cite{Ols98,GraBlu04} 
allowing Feshbach resonance tuning \cite{Rob01} of the effective 1D 
interactions to very large values, such correlations are greatly
enhanced. This has allowed experimental verification 
\cite{Par04Kin04,Kin0506} of the fermionization of
bosonic ultracold vapors in such geometries predicted by the Fermi-Bose (FB) 
mapping method, which was introduced in 1960 \cite{Gir60Gir65} and
used to obtain the exact $N$-particle ground and excited states of a 1D
gas of impenetrable point bosons, the Tonks-Girardeau (TG) gas. 
Apart from recent work of
Imambekov and Demler \cite{ImaDem06}, theoretical work on 1D mixtures
has used approximations which are not valid in the highly-correlated regime.
Exact results in \cite{ImaDem06}, based on the Bethe
ansatz method of Lieb and Liniger (LL) \cite{LieLin63}, are limited to the 
spatially uniform case (no longitudinal trap potential).
We present here exact solutions for several models of mixtures of
strongly correlated 1D systems including that of \cite{ImaDem06}, for both
untrapped and harmonically trapped mixtures.

{\it Bose-Fermi mixture:} We consider a 1D mixture 
of $N_B$ bosons with point hard-core boson-boson interactions and of $N_F$ 
 noninteracting fermions with point hard-core boson-fermion interactions, and we assume equal masses  $m_B=m_F=m$. This model could be realized by choosing a bosonic and fermionic isotope of a given alkali element (e.g.~$^6$Li - $^7$Li).  We indicate the boson and fermion 
coordinates by $X_B=(x_{1B},\cdots,x_{N_B B})$,   
$X_F=(x_{1F},\cdots,x_{N_F F})$. Our treatment includes the case where the mixture is subjected to an external potential (e.g.~a harmonic trap
 $v(x)=\frac{1}{2}m\omega^2 x^2$), provided that the potentials acting on the bosons and on the fermions are the same, i.e.,
$v_B(x_{jB})=v(x_{jB})$ and $v_F(x_{\ell F})=v(x_{\ell F})$.
The Schr\"{o}dinger Hamiltonian is
$\hat{H}=\hat{H}_B+\hat{H}_F+\hat{H}_{BB}+\hat{H}_{BF}$ with
\begin{eqnarray}\label{TGBF}
\hat{H}_B&=&\sum_{j=1}^{N_B}[-(\hbar^2/2m)\partial^2/\partial x_{jB}^2
+v(x_{jB})]\nonumber\\
\hat{H}_F&=&\sum_{\ell=1}^{N_F}[-(\hbar^2/2m)\partial^2/\partial x_{\ell F}^2
+v(x_{\ell F})]\ .
\end{eqnarray}
The impenetrable point BB and BF interactions can be formally represented as
sums of LL delta function interactions \cite{LieLin63} 
$g_{BB}\delta(x_{jB}-x_{\ell B})$ and 
$g_{BF}\delta(x_{jB}-x_{\ell F})$
with coupling constants $g_{BB}\to +\infty$ and 
$g_{BF}\to +\infty$, but they are more conveniently represented by constraints 
that the many-body wavefunction $\Psi(X_B,X_F)$ vanishes at all BB and BF
collision points,
after which $\hat{H}_{BB}$ and $\hat{H}_{BF}$ can be omitted from the
Hamiltonian.

The exact solution can be obtained as follows.
Construct a ``model wavefunction'' $\Psi_M(X_B,X_F)$ which is a Slater
determinant of $N=N_B+N_F$ orthonormal orbitals 
$u_1(x),\cdots,u_{N}(x)$ occupied by all $N$ bosons and fermions, with
all possible permutations of the atoms among the orbitals:
$\Psi_M=\sum_P \varepsilon(P)u_1(Px_1)\cdots u_{N}(Px_{N})$. Here 
$x_1,\cdots,x_{N}$ are 
$x_{1B},\cdots,x_{N_B B},x_{1F},\cdots,x_{N_F F}$, the sum runs over all
$N!$ possible permutations of these variables \emph{including
permutations exchanging bosons with fermions}, and $\varepsilon(P)$ is the
usual $\pm 1$ sign of the permutation. This wave function vanishes not only at 
the points $x_{jF}=x_{\ell F}$
required by fermionic antisymmetry, but also at the points $x_{jB}=x_{\ell B}$
and $x_{jB}=x_{\ell F}$ required by the BB and BF hard core constraints.
Its improper symmetry under BB and
BF exchange is repaired by a modified FB mapping. Define
a mapping function $A(X_B,X_F)$ by 
\begin{equation}\label{map}
A=\prod_{1\le j<\ell\le N_B}\text{sgn}(x_{jB}-x_{\ell B})
\prod_{j=1}^{N_B}\prod_{\ell=1}^{N_F}\text{sgn}(x_{jB}-x_{\ell F})
\end{equation}
where the sign function $\text{sgn}(x)$ is $+1\ (-1)$ if $x>0\ (x<0)$.
Then the physical wavefunction is $\Psi(X_B,X_F)=A(X_B,X_F)\Psi_M(X_B,X_F)$. 
$\Psi_M$ is an exact many-body energy eigenstate if the orbitals 
$u_{\nu}$ are eigenfuctions of the single-particle Schr\"{o}dinger
Hamiltonian $-(\hbar^2/2m)\partial^2/\partial x^2+v(x)$ with eigenvalues
$\epsilon_{\nu}$. Since $A$ is constant except for jumps at nodes
of $\Psi_M$, it follows that the physical state $\Psi$ is an
energy eigenstate with eigenvalue $\sum_{\nu}\epsilon_{\nu}$  
and is symmetric under permutations of bosons and
antisymmetric under permutation of fermions. 
The ground state is a filled Fermi sea of the lowest $N$ 
orbitals and excited states are generated by choosing higher orbitals. 
An important case for experiments is that of harmonic trapping,
$v(x)=\frac{1}{2}m\omega^2 x^2$. Then the ground state is a 
straightforward generalization of that of the 
trapped TG gas \cite{GirWriTri01}. The orbitals are Hermite-Gaussians
$\phi_n(x)=\pi^{-1/4}(2^n n!x_{\text{osc}})^{-1/2}e^{-Q^2/2}H_n(Q)$
with $n=0,1,\cdots,N$, where $Q=x/x_{\text{osc}}$ and
$x_{\text{osc}}=\sqrt{\hbar/m\omega}$. $\Psi_{0M}$ can be reduced to
Bijl-Jastrow form by determinantal algebra as in the original solution
\cite{Gir60Gir65} for the untrapped TG gas and the recent one 
\cite{GirWriTri01} for the trapped TG gas, yielding
$\Psi_{0M}\propto [\prod_{1\le j<\ell<N}(x_j-x_\ell)]
\prod_{j=1}^{N}\phi_0(x_j)$ where
$x_1=x_{1B},\cdots,x_{N_B}=x_{N_B B},x_{N_B+1}=x_{1F},\cdots,
x_{N}=x_{N_F F}$. By applying the mapping function 
$A$ to $\Psi_{0M}$ one finds for the ground state
\begin{eqnarray}\label{TGBFground}
&&\Psi_0=[\prod_{1\le j<\ell\le N_B}|x_{jB}-x_{\ell B}|]
[\prod_{j=1}^{N_B}\prod_{\ell=1}^{N_F}|x_{jB}-x_{\ell F}|]\nonumber\\
&&\times[\prod_{1\le j<\ell\le N_F}(x_{jF}-x_{\ell F})]
\phi_0(x_{1B})\cdots\phi_0(x_{N_B B})\nonumber\\
&&\times\phi_0(x_{1F})\cdots\phi_0(x_{N_F F})\ .
\end{eqnarray}
apart from normalization. The solution for the homogeneous system of
size $L$ with periodic boundary conditions  differs
only by omission of the Gaussian factors $\phi_0$  and by  replacement of each coordinate difference $x-x'$ by
$\sin[\pi(x-x')/L]$ \cite{Gir60Gir65}. 
The bosonic and fermionic reduced one-body density matrices are defined as 
$\rho_{1B}(x,x')=\mathfrak{N}_B\int\Psi_0(x,X_B',X_F)
\Psi_0^*(x',X_B',X_F)dX_B'dX_F$ and 
$\rho_{1F}(x,x')=\mathfrak{N}_F\int\Psi_0(X_B,X_F',x) \Psi_0^*(X_B,X_F',x')
dX_BdX_F'$, where 
$X_B'=(x_{2B},\cdots,x_{N_B B})$ and  $X_F'=(x_{1F},\cdots,x_{N_F-1, F})$, and 
$\mathfrak{N}_B$, $\mathfrak{N}_F$ are normalization constants fixed by
the conditions $\int\rho_{1B}(x,x)dx=N_B$ and  $\int\rho_{1F}(x,x)dx=N_F$. 

{\it Density profiles}: The bosonic and fermionic single-particle densities
$\rho_B(x)\equiv \rho_{1B}(x,x)$ and $\rho_F(x)\equiv\rho_{1F}(x,x)$ are 
the same as those
of $\Psi_{0M}$, since they depend only on $|\Psi_0|^2$, 
and $A^2=1.$ Since $\Psi_{0M}$ is completely antisymmetric under 
permutations of \emph{all} $N$ particles, $\rho_B$ and
$\rho_F$ are both proportional to the
density $\rho_{\text{TG}}$ of a harmonically trapped TG gas of $N$
bosons:
$\rho_B(x)=(N_B/N)\rho_{\text{TG}}(x)$ and 
$\rho_F(x)=(N_F/N)\rho_{\text{TG}}(x)$
where \cite{GirWriTri01} 
$\rho_{\text{TG}}(x)=\sum_{n=0}^{N-1}|\phi_n(x)|^2$. These exact
results contrast strongly with local density approximation  
results in the TG limit
 \cite{ImaDem06}, which show BF phase separation with 
$\rho_B$ more concentrated in the center, although their total density 
$\rho_B(x)+\rho_F(x)$ agrees closely with ours. Note that the exact ground 
state is highly degenerate in the TG limit \cite{Note1} 
due to the fact that there is no required symmetry under BF exchange.
Our choice (\ref{TGBFground}) is the most symmetrical, has hard-core cusps at 
all BB and BF collisions and antisymmetry nodes at all FF collisions, and it 
will be shown below that if the degeneracy is lifted by making the BB and BF 
interactions finite, then (\ref{TGBFground}) corresponds to the lowest member
of the split ground manifold, and hence our $\rho_B(x)$ and $\rho_F(x)$ are the
TG limit of the true ground state partial densities for large but finite 
repulsions. Nevertheless, in the TG limit the density profiles are labile, 
since linear combinations of the degenerate ground basis \cite{Note1} have 
different density profiles.

{\it Density-density correlations and collective 
excitation spectrum:} 
The BB, FF, and BF density-density correlation functions for this model  
depend only on $\Psi_{0M}$, and hence are all the same as that of an ideal 
Fermi gas or TG gas of $N$ particles apart from normalization; see Eqs. (10)
and (11) of \cite{GirWriTri01}. This implies,  
e.g, that  no composite fermions  are found (as is the case for a BF 
mixture on a lattice for particular values of $g_{BF}$ and $g_{BB}$ 
\cite{Lew04}). The spectrum of collective excitations
is given by the poles of the dynamic structure factor, and hence 
coincides with that of an ideal Fermi gas. For the case of 
harmonic trapping the spectrum is given by integer multiples of the trapping 
frequency $\omega$ for both in-phase and out-of-phase modes, 
in  disagreement with \cite{ImaDem06}. 

{\it Momentum distributions}: Momentum 
distributions, Fourier transforms of the reduced
one-body density matrices with respect to $x-x'$,
differ greatly 
between bosons and fermions. Consider first the bosonic one-particle density 
matrix $\rho_{1B}(x,x')$ and momentum distribution $n_B(k)$.
Using (\ref{TGBFground}) one finds that $\rho_{1B}$ 
is proportional to the one-body density matrix of a pure
TG gas of $N$ bosons, 
$\rho_{1B}(x,x')=(N_B/N)\rho_{1TG}(x,x')$, and hence $n_B(k)=(N_B/N)n_{TG}(k)$.
Although $\rho_{1TG}$
is not known analytically, the asymptotics of $n_{TG}(k)$ are known
for a homogeneous system of density $\rho_{TG}$,
and are given by $n_{TG}(k)\propto |k|^{-1/2}$ for $k\ll\rho_{TG}$
\cite{Len64VaiTra79} and by $n_{TG}(k)\propto |k|^{-4}$ for $k \gg\rho_{TG}$ 
\cite{MinVigTos01OlsDun03}.
Determination of $\rho_{1F}(x,x')$ and $n_F(k)$ is much more difficult. 
We detail here the derivation for 
the homogeneous system.
Comparing Eq. (\ref{TGBFground}) with $\Psi_{0}$ of 
the pure TG gas \cite{Gir60Gir65}  
we obtain
$\rho_{1F}(x,x')=\mathfrak{N}_F\int dx_1...dx_{N-1}|\Psi_{0M,N-1}|^2
\prod_{1\le j\le N_B} \text{sgn}(x_j-x) \text{sgn}(x_j-x')
\prod_{1\le j\le N-1} 4 \sin(\pi (x_j-x)/L) \sin(\pi(x_j-x')/L)$
where $\Psi_{0M,N-1}$ is $\Psi_{0M}$ for $N-1$ particles.
The determinantal
expression   
$\Psi_{0M,N-1}=\det[e^{ik_j
x_\ell}]_{j,\ell=1,...N-1}$, where
$k_j=-(N-2)\pi/L,-(N-4)\pi/L,...,(N-4)\pi/L, (N-2)\pi/L$  
and some algebra reduce $\rho_{1F}$ to the sum of a product
of one-dimensional integrals, of which the first $N_B$ involve
$b(t,x,x')=|\sin(\pi(t-x)/L)||\sin(\pi(t-x')/L)|$ as in a TG gas, and the
remaining $N_F$ contain 
$f(t,x,x')=\sin(\pi(t-x)/L)\sin(\pi(t-x')/L)$
as in an ideal Fermi gas. Combining the dissimilar factors by the
``phase trick'' of \cite{ImaDem06,Note3}, one finds the final result 
\begin{equation}
\rho_{1F}(x,x')=\alpha\int \frac{d
  \phi}{2\pi} e^{-iN_B\phi} \det_{j,k=1,...N-1}[ u_{j,k}(x,x',\phi)]
\end{equation}
where $\alpha
=(N_F/NL) {\cal C(N_B,N_F)}$, 
and 
$u_{j,k}(x,x',\phi)=\frac{4}{L}\int_0^L dt [e^{i\phi}{\rm sgn}(t-x)
{\rm sgn}(t-x')+ 1]\sin(\pi(t-x)/L )\sin(\pi(t-x')/L)e^{i 2\pi
  t(j-k)/L}$. 
This can be extended to a harmonic trap.

{\it Ground and excited states for large but finite BB and BF repulsion:} 
Suppose now that the interactions are finite LL interactions \cite{LieLin63}
$g_{BB}\delta(x_{jB}-x_{\ell B})$ and $g_{BF}\delta(x_{jB}-x_{\ell F})$.
Define dimensionless coupling constants 
$\lambda_{BB}=\frac{g_{BB}}{2\hbar}\sqrt{\frac{m}{2\hbar\omega}}$ and
$\lambda_{BF}=\frac{g_{BF}}{2\hbar}\sqrt{\frac{m}{2\hbar\omega}}$. The
TG limit, to which our ground state  (\ref{TGBFground}) corresponds, is 
$\lambda_{BB}\to+\infty$ and $\lambda_{BF}\to+\infty$. As previously noted 
this state is only one member of a degenerate ground manifold \cite{Note1},
and we now examine how this degeneracy is split when $\lambda_{BB}$ and 
$\lambda_{BF}$ are large but finite. 
The exact solutions of the two-body problem are known 
\cite{Bus98} and the energy eigenfunctions
$\Psi_\nu(x_{BF})$ of the relative motion in a harmonic well are
expressible in terms of 
parabolic cylinder functions $D_\nu(\xi)$ \cite{Gra80} where 
$\xi=x_{BF}\sqrt{\frac{m\omega}{\hbar}}$. The corresponding relative 
energy eigenvalues are
$E_\nu=(\nu+\frac{1}{2})\hbar\omega$ where
$\nu$ is a solution of the transcendental equation
$\Gamma(\frac{1}{2}(1-\nu))/\Gamma(-\frac{1}{2}\nu)=-\lambda_{BF}$,  
and reduce to the harmonic oscillator (HO) quantum numbers 
$n=0,1,2,\cdots$ in the noninteracting limit. 
Here we want the opposite
limit $\lambda_{BF}\gg 1$. In the TG limit
all the eigenfunctions are
$\Psi_n(x_{BF})=A(x_{BF})\Psi_{nM}(x_{BF})$ where $A$ is a
mapping function \cite{Note1} and the model functions $\Psi_{nM}$ are the 
\emph{odd} HO functions $u_n$ with $n=1,3,5,\cdots$ 
so as to vanish at contact. 
There are
two possible mapping functions: $A_1=\text{sgn}(x_{BF})$
and $A_2=+1$, and the ground state is
twofold degenerate, with even and odd eigenfunctions 
$\Psi_1^+=\text{sgn}(x_{BF})u_1(x_{BF})$ and 
$\Psi_1^-=A_2u_1=u_1(x_{BF})$ and  energy
$E_1=\frac{3}{2}\hbar\omega$. For large but finite
$\lambda_{BF}$ the lowest even eigenfunction is 
$\Psi_1^+=D_{\nu_1}(|\xi|)$ 
where $\nu_1$ is the solution of the above transcendental equation closest to
$1$.
$\Psi_1^+$ does not vanish at $x_{BF}=0$ and has an LL cusp
\cite{LieLin63}  
there due physically to the delta function interaction
and mathematically to the absolute value in its argument \cite{Note1.1}.
Using gamma function identities and Taylor expansions one finds
$\nu_1=1-\frac{\lambda_{BF}^{-1}}{\sqrt{\pi}}+O(\lambda_{BF}^{-2})$. On the
other hand, the lowest \emph{odd} eigenfunction $\Psi_1^-$ is the lowest 
odd unperturbed HO 
eigenfunction $u_1$, since $u_1\delta(x_{BF})=0$. Hence the symmetric
solution $\Psi_1^+$ is the ground state and $\Psi_1^-=u_1$ is the first excited
state, with excitation energy 
$\Delta E=\frac{\hbar\omega}{\lambda_{BF}\sqrt{\pi}}+O(\lambda_{BF}^{-2})$.
More generally, all of the eigenfunctions in the order of increasing energy are
$D_{\nu_1}(|\xi|),\  u_1,\  D_{\nu_2}(|\xi|),\  u_2,\cdots$,
alternately even and odd with $0,1,2,\cdots$ nodes. This is in agreement with 
Sturm-Liouville theory and the theorem that the ground state 
of a Boltzmann or Bose system is real, nonnegative, and hence nodeless. 

For arbitrary $N_B$ and $N_F$ and finite coupling constants the exact solutions
are known in the untrapped case only for the special case 
$\lambda_{BB}=\lambda_{BF}$. However, the above theorem on 
ground state symmetry has a generalization according to which the
ground state is nodeless \emph{except for Fermi antisymmetry nodes}.
It follows that the ground state vanishes only at FF contact points
$x_{jB}=x_{\ell F}$, and at all BB and BF collision points it has interaction
cusps of LL form \cite{LieLin63}. By comparison with Eq. (\ref{TGBFground}),
one concludes that (\ref{TGBFground}) is the TG limit of the 
finite-interaction ground state; all other choices \cite{Note1}
of the TG mapping $A$ correspond to excited states. The manner
in which the degeneracy of the TG ground manifold splits when 
$\lambda_{BB}$ and $\lambda_{BF}$ are made large but finite is related to 
symmetries of all the different mappings \cite{Note1} yielding the different 
ground states degenerate in the TG-limit. Consider first
a mapping $A^{(0)}$ differing from (\ref{map}) by omission of \emph{all} of the
factors $\text{sgn}(x_{jB}-x_{\ell F)}$. 
The corresponding mapped state is $\Psi_0^{(0)}=A^{(0)}\Psi_{0M}$ where
$\Psi_{0M}$ is the previously-defined fermionic model ground state. $A^{(0)}$
restores the correct symmetry under BB permutations and introduces the
necessary interaction cusps at BB collision points \cite{LieLin63}, 
while leaving $\Psi_0^{(0)}$ antisymmetric not only under FF permutations, but 
also \emph{locally} antisymmetric under BF exchanges, in the sense that
$\Psi_0^{(0)}$ changes sign as $x_{jB}-x_{\ell F}$ passes through zero 
\cite{Note2}.
It follows that $\Psi_0^{(0)}$ is an eigenstate not only of the TG Hamiltonian
$\lambda_{BB}\to\infty$ and $\lambda_{BF}\to\infty$, but also 
of a Hamiltonian with TG BB interactions 
(i.e., $\lambda_{BB}\to\infty$) but \emph{finite} BF interactions,
because the nodes at BF collision points ``kill'' the delta function
interactions, and its energy is the same as that in the TG limit.
It is not the ground state because of the previously-stated 
theorem, and in fact it is the \emph{top} of a ladder of closely-spaced levels 
extending upward from the ground state, with splittings of order 
$\frac{\hbar\omega}{\gamma_{BF}}$ between adjacent levels, obtained by  
restoring more and more collision $\text{sgn}$ factors in the mapping function
and hence collision cusps in the wave function, 
thus generating additional downward shifts of order 
$\frac{\hbar\omega}{\gamma_{BF}}$.

{\it Other soluble models }: In addition to the previously discussed 
Bose-Fermi mixture (which we denote as Model I) we can solve several other 
models with a similar approach. Model II is a mixture of two noninteracting 
Fermi gases with interspecies point hard-core repulsions, and model III is a 
mixture of two Bose gases A and B with inter- and intraspecies hard-core point 
interactions. Model II could be realized, e.g., by choosing two hyperfine 
levels of $^{40}$K or $^6$Li and by tuning to very large and repulsive the 
interspecies interactions by a confinement-induced Feshbach resonance and 
analogously model III could be realized by choosing two hyperfine levels of a 
bosonic alkali atom e.g., Na or Rb. Models I-III all start from the same
model state $\Psi_M$, so we only state final results, again for harmonic 
trapping. For both models II and III the density profiles are proportional to 
those of a TG gas of $N=N_A+N_B$ bosons, and 
in model III the momentum distributions for both
components are also proportional to those of a $N$-particle  TG gas, 
while for model II the $A$-component momentum distribution $n_A(k)$
is equal to $n_F(k)$ for model I with $N_A$ fermions and  $N_B$ 
bosons, and similarly for $n_B(k)$ with $A$ and $B$ interchanged.
These models have degenerate ground states by the same mechanism as model I.

Finally, we can consider binary mixtures with large, attractive, odd-wave 
interactions between the two species due to a p-wave Feshbach resonance, 
in the so-called fermionic Tonks-Girardeau 
(FTG) limit \cite{GraBlu04,GirOls03,GirNguOls04} wherein the fermionic wave 
function is nonzero with sign change at contact. Model IV is a mixture of two 
Bose gases A and B with no AA or BB interactions but an AB interaction of FTG 
form, model V is a mixture of ideal Bose gas B and FTG gas F with FTG BF 
interactions, 
and model VI is a mixture of two FTG gases A and B with FTG AB interactions.
These models all have the same
model state $\Psi_M$,  a mixture of two dissimilar ideal Bose gases, while the 
mapping function $A$ has $\text{sgn}$ factors for all collisions where an
FTG interaction is desired \cite{GraBlu04,GirOls03,GirNguOls04}.
The momentum distributions for the spatially uniform case on a ring
have Lorentzian form as for the FTG gas \cite{BenErkGra05}, with the range of 
the density matrix of each component 
determined by the sum of the densities of all 
components (either the same or different) with which it interacts. 

We have identified further models with attractive interspecies FTG 
interactions solvable by mapping from ideal gas mixtures to
mixtures of an ideal Fermi gas or TG Bose gas and an ideal Bose gas or FTG 
Fermi gas, or a mixture of two dissimilar ideal Fermi gases or TG Bose gases, 
and we believe that this covers all two-component mixture models solvable by 
mapping functions everywhere $\pm 1$, with jumps at contact points which 
convert Fermi antisymmetry nodes to Bose hard core cusps and Bose-Bose or 
dissimilar species contact points to infinite FTG attraction discontinuities. 

{\it Experimental realizability and applications:} Since the TG regime in pure 
ultracold gases has been reached \cite{Par04Kin04,Kin0506} and p-wave 
resonances required for the FTG regime have been observed
\cite{RegTicBohJin03}, several of these strongly 
interacting mixture models should be experimentally realizable. 
The low-lying level structure we have found for Bose-Fermi mixtures 
might be exploited for quantum computation.
\begin{acknowledgments}
This work was initiated at the Institut Henri Poincar\'{e}-Centre Emile Borel 
(IHP) in  Paris, during the 2007 workshop ``Quantum Gases''. We  
are grateful to the organizers, to the IHP 
for hospitality and support, and to Jason Ho and Gora Shlyapnikov  
for helpful suggestions. We are also very grateful
to Hrvoje Buljan for pointing out to us the degeneracy of the TG Bose-Fermi
ground state. Research of M.D.G. is partially 
supported by the U.S. Office of Naval Research, and that of A.M. by  the Centre
National de la Recherche Scientifique (CNRS).
\end{acknowledgments}

\begin{thebibliography}{22}
\bibitem{Note1.1} The $D_\nu(y)$ are neither even nor odd , being sums of even 
and odd hypergeometric
functions \cite{Gra80}. They vanish like $y^\nu e^{-y^2/4}$ as 
$y\to +\infty$ and diverge like $|y|^{-\nu-1}e^{y^2/4}$ as $y\to -\infty$,
but our even solution depends only on $\xi=|y|$.
\end{thebibliography}
\end{document}